\documentclass[runningheads]{llncs}
\usepackage{graphicx}
\usepackage{amsmath}
\usepackage{amssymb}
\usepackage[caption=false]{subfig}
\usepackage{booktabs}
\usepackage{stmaryrd}

\begin{document}
\title{Contour-based Bone Axis Detection for X-Ray Guided Surgery on the Knee\thanks{The authors gratefully acknowledge funding of the Erlangen Graduate School in Advanced Optical Technologies (SAOT) by the Bavarian State Ministry for Science and Art.}}
\titlerunning{Contour-based Bone Axis Detection for X-Ray Guided Surgery}
\author{Florian Kordon\inst{1,2,3}\orcidID{0000-0003-1240-5809} \and
Andreas Maier\inst{1,2}\orcidID{0000-0002-9550-5284} \and
Benedict Swartman\inst{4} \and
Maxim Privalov\inst{4} \and
Jan Siad El Barbari\inst{4} \and
Holger Kunze\inst{3}\orcidID{0000-0002-7021-2370}}
%index{Kordon, Florian}
%index{Maier, Andreas}
%index{Swartman, Benedict}
%index{Privalov, Maxim}
%index{El Barbari, Jan Siad}
%index{Kunze, Holger}

% 
% 
\authorrunning{Kordon et al.}
\institute{
Pattern Recognition Lab, Friedrich-Alexander-Universität Erlangen-Nürnberg (FAU), Erlangen, Germany \\
\email{florian.kordon@fau.de}
\and
Erlangen Graduate School in Advanced Optical Technologies (SAOT), Friedrich-Alexander-Universität Erlangen-Nürnberg (FAU), Erlangen, Germany 
\and
Siemens Healthcare GmbH, Forchheim, Germany
\and 
Department for Trauma and Orthopaedic Surgery, BG Trauma Center Ludwigshafen, Ludwigshafen, Germany}
\maketitle              % typeset the header of the contribution
\begin{abstract}
The anatomical axis of long bones is an important reference line for guiding fracture reduction and assisting in the correct placement of guide pins, screws, and implants in orthopedics and trauma surgery. This study investigates an automatic approach for detection of such axes on X-ray images based on the segmentation contour of the bone. For this purpose, we use the medically established two-line method and translate it into a learning-based approach. The proposed method is evaluated on 38 clinical test images of the femoral and tibial bone and achieves a median angulation error of $0.19^{\circ}$ and $0.33^{\circ}$ respectively. An inter-rater study with three trauma surgery experts confirms reliability of the method and recommends further clinical application.

\keywords{Surgical Planning \and Orthopedics \and Bone Axis Detection \and X-Ray Imaging \and Intra-operative Guidance}
\end{abstract}
\section{Introduction}
\begin{figure}[tb]
    \centering
    \subfloat[Palmar tilt and radial inclination angle on the wrist joint \cite{Dee.2000,Kreder.1996,Schmitt.2015,Watson.2016}. \label{fig1a}]{
       \makebox[0.45\textwidth][c]{\includegraphics[height=0.38\textwidth]{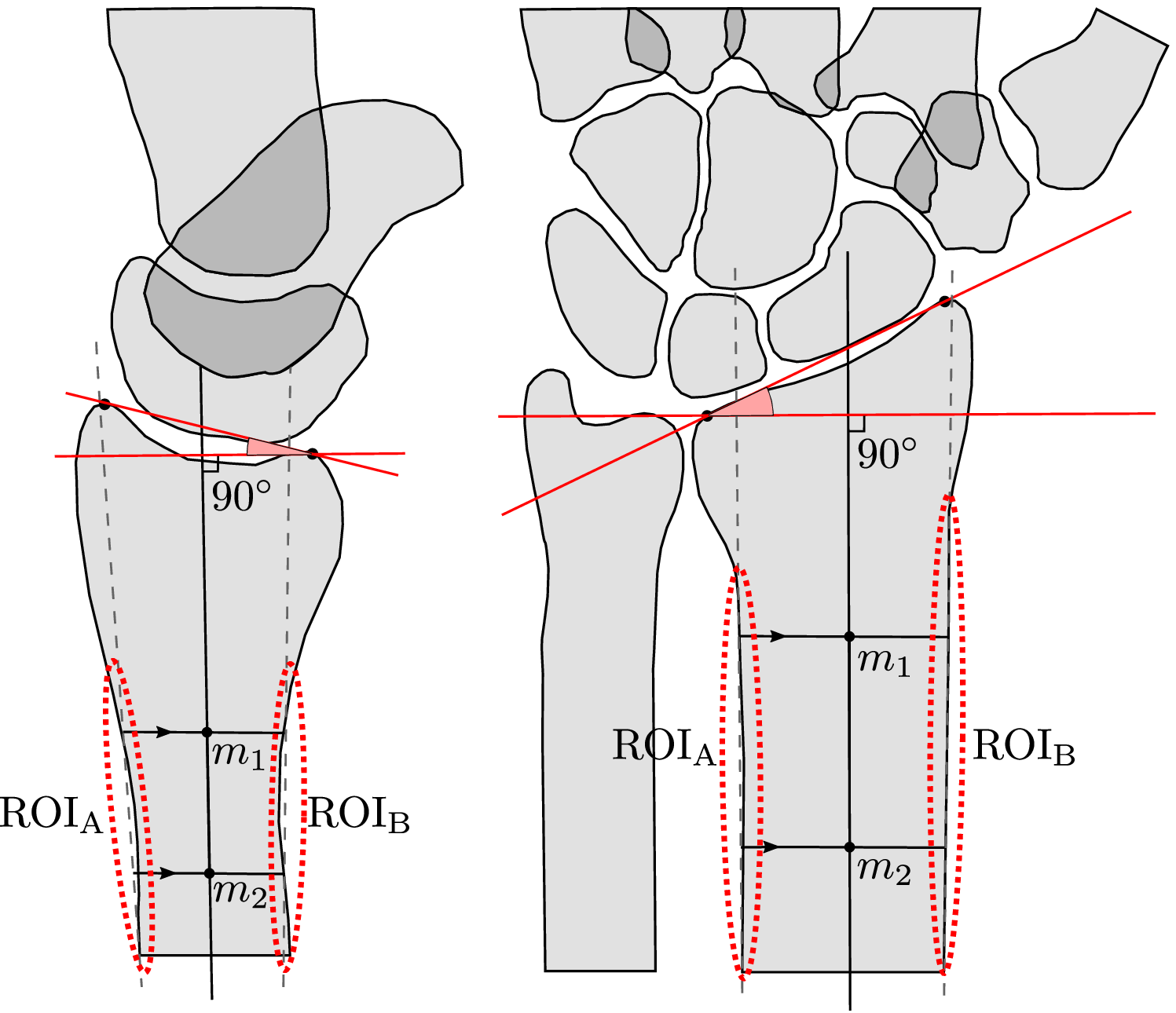}}
    }\quad
\subfloat[Baumann angle on frontal radiograph of the elbow \cite{Silva.2010,Williamson.1992}. \label{fig1c}]{
    \makebox[0.45\textwidth][c]{\includegraphics[height=0.38\textwidth]{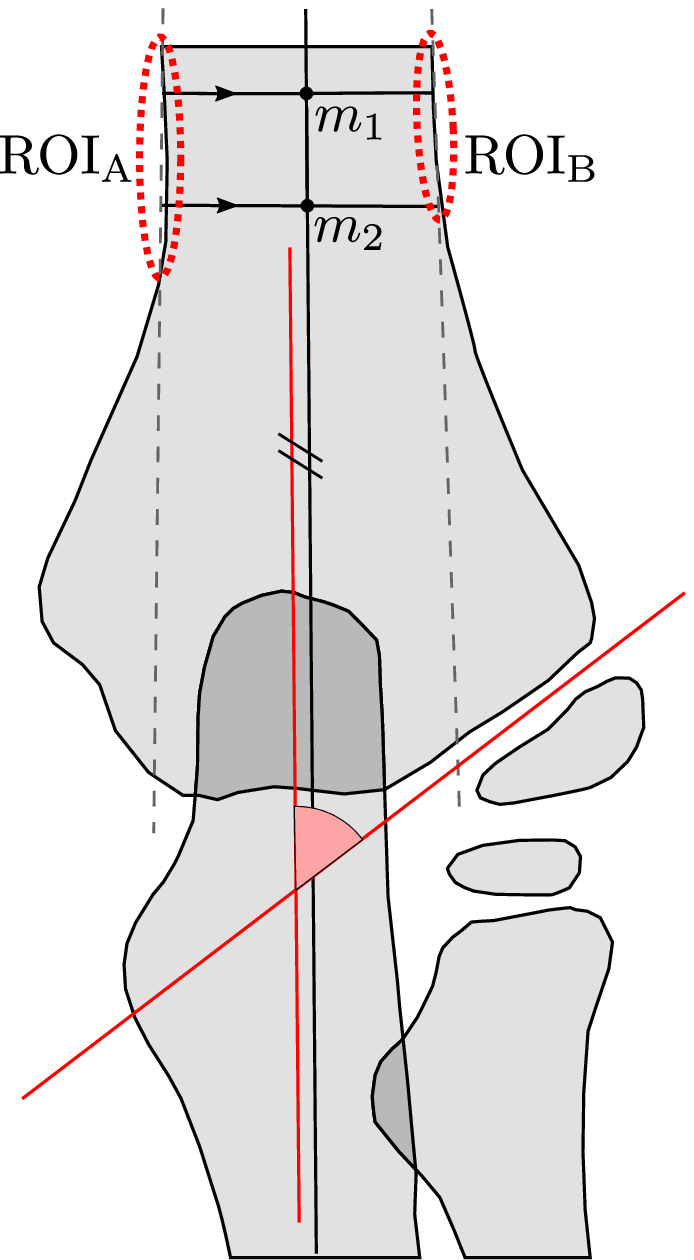}}
} \\
\subfloat[Angles for tibial intramedullary nail insertion (solid) \cite{Franke.2018} and transtibial tunnel drilling (dashed) \cite{Hiesterman.2011,Johannsen.2013}. \label{fig1b}]{
    \makebox[0.45\textwidth][c]{\includegraphics[height=0.38\textwidth]{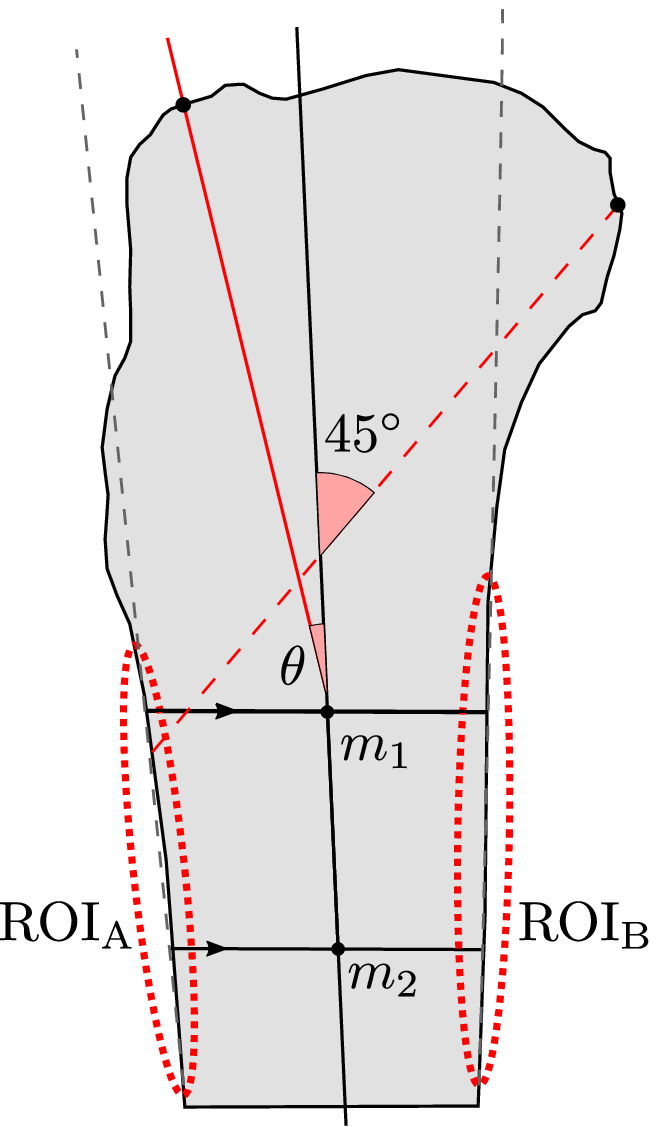}}
}\quad
\subfloat[Approximation of knee flexion angle based on femoral and tibial bone axes. \label{fig1d}]{
    \makebox[0.45\textwidth][c]{\includegraphics[height=0.38\textwidth]{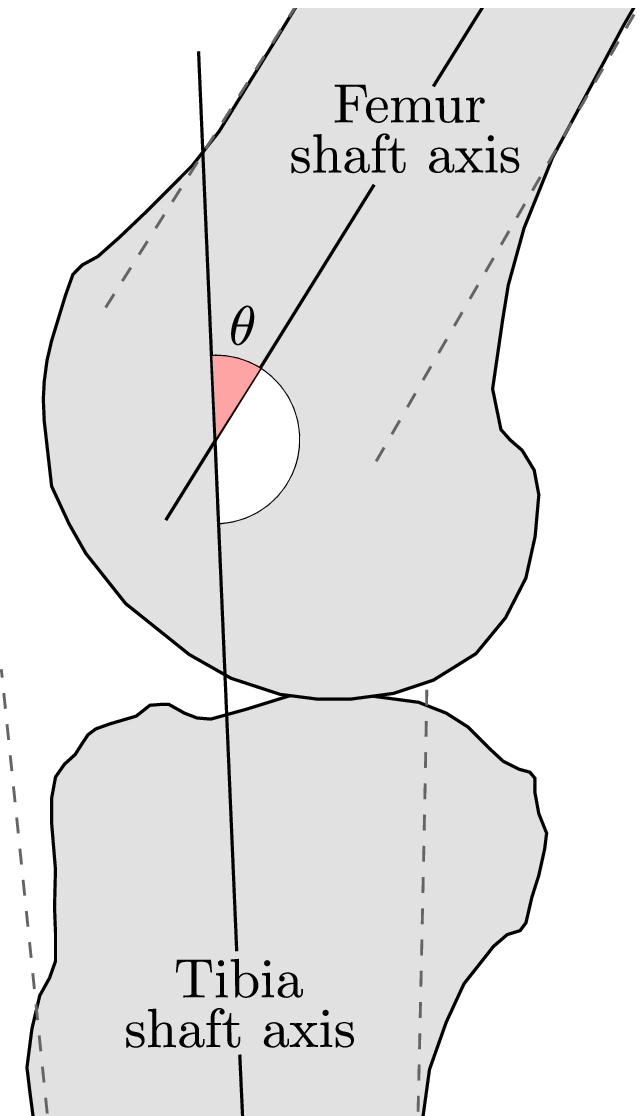}}
}
\caption{Examples for using the shaft axis of long bones as reference line.} 
\label{fig1}
\end{figure}

The reconstruction of anatomical joint surface and angular relationships is a paramount aspect in surgical management of fractures or ligament injuries. Intra-operative fluoroscopic guidance, 3D imaging, or navigation is typically used to ensure anatomically and mechanically correct reduction, so that irregular joint loading and complications caused by aberrant biomechanics can be alleviated or avoided. Moreover, for technically demanding procedures, a pre-operative planning sketch is obligatory and helps the surgeon to achieve operational safety \cite{Ewerbeck.2014}. In many of these planning and verification steps, the bone axis serves as an important reference line (Fig.~\ref{fig1}). While planning such axes can be easily done on pre-operative static data, doing so consistently on live images during surgery is inherently more complex due to motion and a limited field of view. In addition, non-sterile interaction with a planning software is unwanted. For this reason, axial alignment is typically verified by visual inspection and use of hardware-based solutions such as the cable method, alignment rods, goniometers, or optical navigation amongst others \cite{Krettek.1998,Lee.2018,Waelkens.2016}. However, these methods either increase task complexity, are inherently imprecise, or require an open reduction or additional incisions regardless of the surgical technique used. 
 % 0.392
To this end, several methods were proposed to automate detection of the bone axis on image data. Tian et al. \cite{Tian.2003} compute the femoral shaft axis by using a combination of contour extraction and analysis of intersecting line normals to the shaft contour. They recover the contour by using Canny edge detection and identify the relevant straight line sections with Hough transformation and active contour mode via Gradient Vector Flow. While this approach can deal with truncated bones, it prerequisites the bone to be oriented in an upward position on the X-ray image to isolate the relevant intersection points. Donnelley et al. \cite{Donnelley.2008} use a scale-space approach and approximate line straight parameters via Hough transformation. To deal with ambiguous peak spread in the dual space encountered in real-world radiographs, this methods relies on prior spread quantification which falls short in the case of truncated bones. Subburaj et al. \cite{Subburaj.2010} use a 3D-reconstructed bone model from pre-operative CT scans. They combine geometrically detected landmarks and maximal inscribed sphere fitting to detect the medial axis, which is then used for identification of anatomical and mechanical axes. Although very accurate results can be achieved, such 3D information is oftentimes not available and requires registration with the intra-operative 2D image. 

To circumvent these limitations, we propose a simple and clinically motivated image-guided approach for detection of the anatomical axis of long bones on 2D X-ray images. We translate the established two-line/two-circle manual method \cite{Hiesterman.2011,James.2014,Johannsen.2013,Kostogiannis.2011} to a learning based extraction of anatomical features and subsequent geometric construction based on segmentation of the bone cortex outline. With reference to \cite{Kordon.2019}, region of interest (ROI) encoding of the relevant contour sections is used to cope with variability in image truncation and arbitrary image rotation. Moreover, the segmentation results can directly be used for registration of the detected axis on fluoroscopic live images. The method is evaluated for the femur and tibia in the knee joint, which are amongst the most prominent anatomies treated in trauma surgery. The reliability of the proposed method is evaluated and confirmed in an inter-rater study with three expert trauma surgeons. 

\section{Methods}
The anatomical axis of long bones in a 2D image plane can be described by two auxiliary lines that follow the orientation of the anterior/posterior or medial/lateral contour of the bone shaft. In contrast to conventional radiographs with rather standardized imaging, this shaft area is usually truncated on intra-operative images due to a limited field of view and a joint-centered acquisition protocol. Furthermore, the largely linear shaft contour can suffer from structural changes due to e.g. bony proliferation. To this end, first the relevant contour sections are estimated and extracted from the image. Subsequently, these sections are masked based on positional probability and smoothed to reduce the influence of outliers. Lastly, the clinically motivated two-line method is used to calculate the bone axis.

\subsection{Likelihood Encoding of Relevant Contour Regions}
\begin{figure}[tb]
    \centering
    \subfloat[Masking of segmentation contour $\mathcal{K}$ by evaluation the predicted likelihood in $\mathrm{ROI}$ and parametrization of the line segment. \label{fig2a}]{
       \makebox[0.45\textwidth][c]{\includegraphics[height=0.318\textwidth]{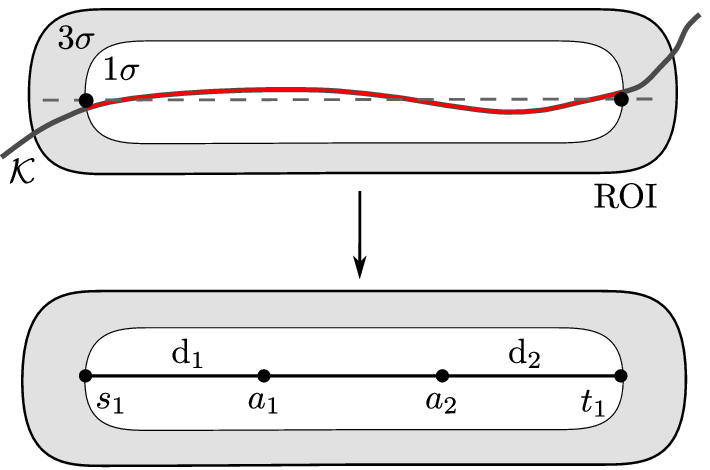}}
    }\quad
\subfloat[Vector projection of intermediary line points onto the opposing segment and geometric construction of the axis points $m_1$ and $m_2$. \label{fig2b}]{
    \makebox[0.45\textwidth][c]{\includegraphics[height=0.343\textwidth]{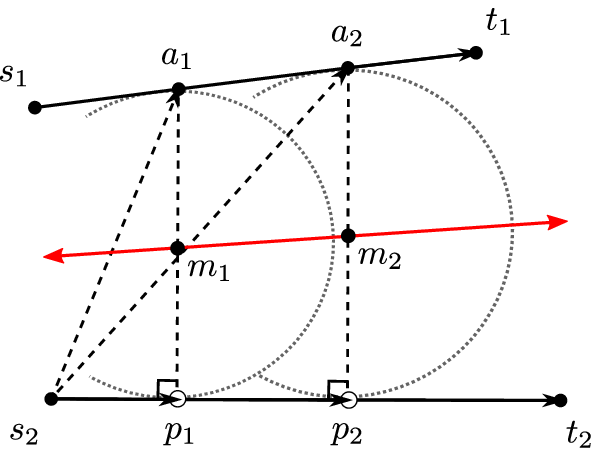}}
}
\caption{Implementation of the two line method for bone axis estimation based on the extracted segmentation contour.}
\label{fig2}
\end{figure}

Given a binary bone segmentation mask $S$ we extract the complete cortex contour $\mathcal{K}$ by using a morphological erosion operation. With a cross-shaped $3 \times 3$ structuring element $X = \{(-1,0),(0,-1),(0,0),(0,1),(1,0)\}$ this equates to 
\begin{equation}
    \mathcal{K} = \mathrm{XOR}\left(S,\,\mathrm{erode}(S,X) \right).
\end{equation}
To constrain the relevant contour section, a ROI similar to \cite{Kordon.2019} is constructed (Fig.~\ref{fig1}). Its bounds are defined by the start and end points of an additional line segment. Positional variance both in the parallel as well as in the orthogonal direction to this line segment is encoded by a 2D Gaussian distribution with a standard deviation of $\sigma=6\,\mathrm{px}$ and truncation bounds at $3\sigma$. This gives us a symmetrical fall-off in probability orthogonal to the line within a margin of $37\,\mathrm{px}$. This spatial likelihood distribution is used to decide whether a contour point should be considered part of the relevant contour region. Since we can assume a mainly linear contour, we argue that using a threshold at $1\sigma$ retains the most probable points while eliminating most outliers (Fig.~\ref{fig2a}). 
% 0.318 / 0.343
\subsection{Axis Construction with Two-Line Method}
The auxiliary contour extension lines are obtained by fitting two linear functions to the pair of relevant contour regions. Since we cannot assume a designated dependent variable due to unknown image rotation, major axis regression\footnote[1]{Ordinary least squares with dependent variable of highest variance is also possible.} is used \cite{Warton.2006}. Given these two lines, we can now perform a geometric construction of the in-between axis based on the midpoints of two parallel and intersecting line segments. This method is known as the two-line method and is a clinically known and trusted procedure especially in pre-operative manual planning \cite{Hiesterman.2011,James.2014,Johannsen.2013,Kostogiannis.2011}. First, a line segment is parametrized for each contour line which is bounded by the relevant contour region. One of these segments is subdivided by two points at distance $\mathrm{d}_1$ and $\mathrm{d_2}$ from the respective start and end points (Fig.~\ref{fig2a}). The actual distances can be selected depending on the target anatomical structure to facilitate easier correction by the user. In a second step, these intermediary points are then projected onto the opposing line segment (Fig.~\ref{fig2b}). This procedure allows for different orientation and length of each segment and is close to clinical practice.

\subsection{Neural Network Architecture}
The proposed construction relies on a segmentation mask of the target long bone and ROI encodings for both relevant contour regions. For combined prediction we use a multi-task variant of the hourglass network architecture by Newell et al. \cite{Kordon.2019,Newell.2016}. This network architecture allows to optimize a joint representation of both tasks and benefits execution time and computational footprint upon inference. We separate segmentation and prediction of ROIs into two tasks. The segmentation task is trained with binary cross entropy to delineate the target bone (foreground) and all other image content (background). The ROIs are optimized by direct matching of the pixel intensity values with a mean squared error loss. In addition, we employ gradient normalization \cite{Chen.2018,Kordon.2019} to cope with different loss function characteristics and task difficulties. To limit the hardware requirements in consideration of the intra-operative application area, we refrain from a stacked network variant.

\subsection{Data and Evaluation}
Network training and geometric construction were evaluated for the femur and tibia on a dataset of 221 clinical X-ray images of the knee joint. Each image was acquired as a lateral standard projection where the outlines of both femoral condyles are aligned. The ground truth segmentation masks and line segments representing the ROIs were annotated by a medically-trained engineer with the \textit{labelme} annotation tool \cite{Kentaro.2016}. Our experiment and evaluation setup on this dataset was two-fold.
\begin{enumerate}
    \item Training of the network in three configurations (a) femur only, (b) tibia only, (c) joint training of femur and tibia, followed by a quantitative evaluation of the performance. For variant (c) the number of output channels for each task head of the network was increased accordingly. 
    \item Assessment of clinical reliability of the automatic axis detection in an inter-rater study. To this end, three expert trauma surgeons (one site) were asked to annotate the femoral and tibial axes on all 38 evaluation images via two axis control points.  
\end{enumerate}
For both experiment series a hold-out test set of 38 images with a $3\,\mathrm{mm}$ calibration sphere was defined. Representative variability in bone truncation and absolute joint rotation was confirmed. The remaining data was split into training and validation subsets of 167/16 images respectively. The data was split up in such a way that disjoint patient groups are ensured in the training/validation and test data sets. Optimization for the first experiment step was performed using Stochastic Gradient Descent (SGD) with a batch size of 2 over 300 epochs on a NVIDIA TITAN RTX graphics card in the PyTorch (v.1.2) Deep Learning framework. We used a learning rate of $2.5e-4$ which we halved every 50 epochs. To aid generalization and to prevent early overfitting, we applied L2 weight decay with a factor of $5e-5$ and a basic online augmentation sequence during training. This sequence comprised affine transformations (scaling, rotation, shearing, horizontal flipping) and margin crops of random strength. Upon propagation in the network, min-max normalization to the interval of $[0, 1]$ was applied and the image resolution was standardized to $256 \times 256\,\mathrm{px}$ by resizing and a subsequent center-crop. All reported results are based on the respective model parameters for which the minimum combined task error on the validation split was observed.

\section{Results and Discussion}

\subsubsection{Bone Segmentation}
The results for bone segmentation by the multi-task neural network variants are given in Table~\ref{tab1}. In general we observe segmentation results which closely resemble the annotated ground truth. Despite missing annotations of other bony structures in the knee joint, the single-anatomy model is capable of delineating the target bone from other structures, even in ambiguous overlap areas. On the other hand, prediction quality of the combined variant does not suffer from a doubling of inference tasks which benefits fast execution time and a smaller computational footprint. A very low contour error indicates that the networks do not only learn the global shape but also successfully capture small details which are often caused by bony erosion and proliferation. This allows for marginal error propagation into geometric axis construction. Segmentation outliers indicated by higher Hausdorff error points are exclusively caused by the inserted measuring spheres which are not represented in the training data.

\begin{table}[b]
\caption{Evaluation of segmentation performance for the femur and tibia (DICE=Sørensen–Dice coefficient; ASD=Average Surface Distance; HD=Hausdorff Distance).}
\label{tab1}
\centering
\begin{tabular}{@{\extracolsep{8pt}}llccccc@{}}
\toprule
        &           & DICE              &  & ASD (mm)           &  & HD (mm) \\
Bone    & Network   & Mean $\pm$ Std    &  & Mean $\pm$ Std     &  & Mean $\pm$ Std \\ \midrule
Femur   & Single    & 0.99 $\pm$ 0.003  &  & 0.57 $\pm$ 0.45    &  & 7.91 $\pm$ 13.86   \\
        & Comb.     & 0.99 $\pm$ 0.004  &  & 0.57 $\pm$ 0.58    &  & 11.38 $\pm$ 22.17 \\ \midrule
Tibia   & Single    & 0.99 $\pm$ 0.005  &  & 0.62 $\pm$ 0.53    &  & 7.07 $\pm$ 7.08 \\
        & Comb.     & 0.99 $\pm$ 0.003  &  & 0.51 $\pm$ 0.16    &  & 4.23 $\pm$ 1.95 \\ \bottomrule
\end{tabular}
\end{table}

\begin{table}[t]
\caption{Angulation and displacement error for the femur and tibia in single/combined anatomy training. Results are reported for the anterior/posterior auxiliary lines and bone shaft axis. The displacement error is constructed as the mean orthogonal point-to-line distance of predicted points $s_1/t_1$, $s_2/t_2$, $m_1/m_2$ onto the respective ground truth axis and combines translation and angulation error components. The best results for each axis are marked in bold ($\mathrm{CI}_{95}$ = 95\% confidence interval).}
\label{tab2}
\centering
\subfloat[Femoral axes detection. \label{tab2a}]{
\begin{tabular}{@{\extracolsep{0.5pt}}llccccccc@{}}
\toprule
 &  & \multicolumn{3}{c}{Angulation (deg)} &  & \multicolumn{3}{c}{Displacement (mm)} \\ \cmidrule(lr){3-5} \cmidrule(l){7-9} 
Axis & Netw. & Mean $\pm$ Std & Median \& $\mathrm{CI}_{95}$ & Max &  & Mean $\pm$ Std & Median \& $\mathrm{CI}_{95}$ & Max \\ \midrule
Ant. & Single & $0.55 \pm 0.98$ & $\mathbf{0.31}$ {[}0.18, 0.57{]} & 6.17 &  & $0.57 \pm 1.25$ & 0.34 {[}0.25, 0.49{]} & 8.09 \\
 & Comb. & $\mathbf{0.48} \pm 0.64$ & 0.34 {[}0.24, 0.52{]} & $\mathbf{3.89}$ &  & $\mathbf{0.47} \pm 0.79$ & $\mathbf{0.31}$ {[}0.19, 0.44{]} & $\mathbf{5.10}$ \\ \midrule
Post. & Single & $\mathbf{0.56} \pm 0.38$ & $\mathbf{0.49}$ {[}0.36, 0.68{]} & $\mathbf{1.50}$ &  & $\mathbf{0.54} \pm 0.33$ & $\mathbf{0.43}$ {[}0.36, 0.55{]} & $\mathbf{1.53}$ \\
 & Comb. & $0.64 \pm 0.44$ & 0.54 {[}0.38, 0.83{]} & 1.94 &  & $0.59 \pm 0.47$ & 0.45 {[}0.36, 0.56{]} & 2.71 \\ \midrule
Shaft & Single & $0.35 \pm 0.42$ & 0.21 {[}0.14, 0.36{]} & 2.54 &  & $0.18 \pm 0.22$ & $\mathbf{0.13}$ {[}0.09, 0.16{]} & 4.11 \\
 & Comb. & $\mathbf{0.28} \pm 0.24$ & $\mathbf{0.19}$ {[}0.11, 0.23{]} & $\mathbf{0.98}$ &  & $\mathbf{0.15} \pm 0.13$ & $\mathbf{0.13}$ {[}0.07, 0.15{]} & $\mathbf{2.65}$ \\ \bottomrule
\end{tabular}
}\\
\subfloat[Tibial axes detection. \label{tab2b}]{
\begin{tabular}{@{\extracolsep{0.5pt}}llccccccc@{}}
\toprule
 &  & \multicolumn{3}{c}{Angulation (deg)} &  & \multicolumn{3}{c}{Displacement (mm)} \\ \cmidrule(lr){3-5} \cmidrule(l){7-9} 
Axis & Netw. & Mean $\pm$ Std & Median \& $\mathrm{CI}_{95}$ & Max &  & Mean $\pm$ Std & Median \& $\mathrm{CI}_{95}$ & Max \\ \midrule
Ant. & Single & $1.59 \pm 1.99$ & $\mathbf{0.64}$ {[}0.43, 1.59{]} & 7.37 &  & $0.86 \pm 0.95$ & $\mathbf{0.48}$ {[}0.35, 0.92{]} & 5.26 \\
 & Comb. & $\mathbf{1.43} \pm 1.62$ & 0.81 {[}0.67, 1.18{]} & $\mathbf{6.98}$ &  & $\mathbf{0.75} \pm 0.50$ & 0.62 {[}0.52, 0.74{]} & $\mathbf{1.93}$ \\ \midrule
Post. & Single & $\mathbf{0.66} \pm 0.55$ & $\mathbf{0.51}$ {[}0.32, 0.68{]} & $\mathbf{1.98}$ &  & $\mathbf{0.38} \pm 0.21$ & 0.36 {[}0.25, 0.45{]} & $\mathbf{0.95}$ \\
 & Comb. & $0.83 \pm 0.84$ & 0.52 {[}0.31, 0.81{]} & 3.37 &  & $0.43 \pm 0.28$ & $\mathbf{0.33}$ {[}0.26, 0.46{]} & 1.30 \\ \midrule
Shaft & Single & $0.78 \pm 0.95$ & 0.48 {[}0.32, 0.78{]} & 4.11 &  & $0.21 \pm 0.19$ & $\mathbf{0.16}$ {[}0.11, 0.25{]} & 0.96 \\
 & Comb. & $\mathbf{0.62} \pm 0.66$ & $\mathbf{0.33}$ {[}0.23, 0.69{]} & $\mathbf{2.65}$ &  & $\mathbf{0.17} \pm 0.11$ & 0.17 {[}0.11, 0.21{]} & $\mathbf{0.38}$ \\ \bottomrule
\end{tabular}
}
\end{table}
\subsubsection{Axes Detection}
The performance of the proposed geometric axis construction is presented in Table~\ref{tab2}. We observe an average angulation error of less than $0.65^{\circ}$ for the anterior and posterior auxiliary lines on both bones and only minor differences between single and combined training. This indicates that the predicted ROIs can provide masking of relevant contour sections on a sufficiently fine scale. We can also qualitatively confirm that the likelihood distribution follows along the actual anatomical contour, albeit this area is only approximated by a straight line in the ground truth annotations. These observations strengthen our assumption that we can retain all relevant contour points by masking at a likelihood threshold of $1\sigma$. In addition, low values for the displacement error (Tab.~\ref{tab2}) indicate minor deviation of the line's shift off the ground truth bone contour. The constructed bone axes generally benefit from the combined training variant and exhibit a comparatively lower maximum error bound (Tab.~\ref{tab2}). Furthermore, it can be observed that by training both anatomies together, the respective confidence intervals taper off and follow the downward shift of the position measure. Based on these results, we chose the combined network for evaluation in the inter-rater study.

\begin{table}[t]
\caption{Comparison of automatically detected shaft axis (Auto) to the annotation of three expert readers (E-1, E-2, E-3) and assessment of inter-rater variability. Due to missing midpoints $m_1$ and $m_2$ in the expert reader annotations, the respective displacement error is based on the two annotated control points. Here, $\mapsto$  denotes a mapping of the $1^{\mathrm{st}}$ rater's control points on the predicted axis of the $2^{\mathrm{nd}}$ rater. $\mapsfrom$ marks a mapping in reverse order.}
\label{tab3}
\begin{tabular}{@{\extracolsep{2pt}}llccccc@{}}
\toprule
 &  & \multicolumn{2}{c}{Femur} &  & \multicolumn{2}{c}{Tibia} \\ \cmidrule(lr){3-4} \cmidrule(l){6-7} 
\begin{tabular}[c]{@{}l@{}}$1^{\mathrm{st}}$ \\ rater\end{tabular} & \begin{tabular}[c]{@{}l@{}}$2^{\mathrm{nd}}$  \\ rater\end{tabular} & \begin{tabular}[c]{@{}c@{}}Angulation\\ (deg)\end{tabular} & \begin{tabular}[c]{@{}c@{}}Displacement\\ (mm)\end{tabular} &  & \begin{tabular}[c]{@{}c@{}}Angulation\\ (deg)\end{tabular} & \begin{tabular}[c]{@{}c@{}}Displacement\\ (mm)\end{tabular} \\ \midrule
Auto & E-1 & 0.61 {[}0.38, 1.01{]} & 0.91 {[}0.57, 1.16{]} &  & 1.48 {[}1.12, 2.23{]} & 1.90 {[}1.36, 2.62{]} \\
Auto & E-2 & 1.12 {[}0.63, 1.82{]} & 0.58 {[}0.39, 0.72{]} &  & 2.36 {[}1.68, 3.07{]} & 1.68 {[}1.22, 2.13{]} \\
Auto & E-3 & 0.49 {[}0.29, 0.68{]} & 0.74 {[}0.43, 1.17{]} &  & 5.76 {[}4.87, 6.44{]} & 3.89 {[}3.38, 4.56{]} \\
 &  &  &  &  &  &  \\
E-1 & E-2 & 0.93 {[}0.39, 1.32{]} & $\mapsto$ 1.85 {[}1.51, 2.21{]} &  & 1.75 {[}1.01, 2.21{]} & $\mapsto$ 1.93 {[}1.60, 2.58{]} \\
 &  &  & $\mapsfrom$ 1.68 {[}1.52, 1.93{]} &  &  & $\mapsfrom$ 2.02 {[}1.72, 2.61{]} \\
E-1 & E-3 & 0.70 {[}0.49, 1.06{]} & $\mapsto$ 1.38 {[}1.20, 1.77{]} &  & 4.53 {[}3.47, 5.25{]} & $\mapsto$ 4.64 {[}3.72, 5.22{]} \\
 &  &  & $\mapsfrom$ 1.82 {[}1.35, 2.15{]} &  &  & $\mapsfrom$ 4.09 {[}3.27, 4.81{]} \\
E-2 & E-3 & 1.02 {[}0.69, 1.29{]} & $\mapsto$ 1.41 {[}1.15, 1.74{]} &  & 3.22 {[}2.40, 4.24{]} & $\mapsto$ 4.71 {[}3.33, 5.47{]} \\
 &  &  & $\mapsfrom$ 1.38 {[}1.20, 1.77{]} &  &  & $\mapsfrom$ 3.71 {[}2.85, 4.25{]} \\ \bottomrule
\end{tabular}
\end{table}
\subsubsection{Inter-rater Comparison}
The reliability of our method in comparison to expert rater annotations is analyzed in Table \ref{tab3}. For the femur, low angulation and displacement errors indicate reliable axis estimates which are independent of the amount of truncation and rotation present in the image data. A significantly higher angular deviation of the tibial axis can be explained by comparatively more divergent contour lines. Together with structural variation of the anterior tibia (tibial tuberosity), this leads to higher complexity and differences in the individual approach to manual annotation. This reasoning is strengthened by comparison with rater E-3 for whom a systematically more posterior position and orientation can be observed. If compared to the differences between expert raters (Tab.~\ref{tab3}), the automatic approach yields very comparable performance and achieves axis predictions that lie within the inter-rater error bounds. It should be noted that agreement between raters could be further increased if a dedicated tool for semi-automatic two-line planning is used. 

\section{Conclusion}
This study investigated a method for automatic detection of the shaft axis on long bone X-rays. The experiments reveal encouraging results which match expert rater performance. A major strength of the proposed method is the flexibility of ROI masking which we use to select relevant sections of the bone contour without strong prerequisites on image truncation and rotation. We see limitations in that no evaluation was performed for bones that suffer from increased antecurvation/recurvation (e.g. due to natural deformity or increased weight bearing) or major occlusion of the contour by surgical implants. In addition, future work should analyze potential extensions to our method to promote axis estimation in cases of multi-fragment fractures. 

\subsubsection{Disclaimer}
The methods and information presented here are based on research and are not commercially available.

\bibliographystyle{splncs04}
\bibliography{paper1514}
\end{document}